\documentclass[useAMS,usenatbib]{mn2e}
\usepackage{graphics}
\usepackage{epsfig}
\usepackage{color}
\usepackage{times}
\usepackage[latin1]{inputenc}

\newcommand{\msun}{M$_\odot$}

\title[Evolution of the galaxy merger rate]{Evolution of the galaxy merger rate in model universes}

\author[A. Mateus]{Ab{{\'\i}}lio Mateus\thanks{E-mail: abilio.mateus@oamp.fr}\\
Laboratoire d'Astrophysique de Marseille, CNRS UMR6110, Traverse du Siphon, 13012 Marseille, France.}

\begin{document}

\pagerange{\pageref{firstpage}--\pageref{lastpage}} \pubyear{2008}

\maketitle

\label{firstpage}

\begin{abstract}
We investigate the evolution of the galaxy merger rate predicted by two semi-analytical galaxy formation models implemented on the Millennium Simulation of dark matter structure growth.  The fraction of merging galaxy pairs at each time-step of the simulation is derived from the galaxy catalogues obtained by the models and the results are compared with various observational estimates of merger fractions taken from the literature. We find a good match between the pair fractions derived from the simulation and the observed counting of galaxy pairs obtained by different sources in the redshift range $0 < z < 1.2$. The predicted evolution of the number of galaxy mergers per Gyr grows with redshift as an exponential rate given by $\Gamma_{mrg} \propto (1+z)^{m}$, with $m$ ranging from $0.6$ to $0.8$ for $0 < z < 2$, depending on the luminosity and mass ratios of the merging galaxies. Our results are in agreement with recent observational results that argue for a flat evolution in the fraction of galaxy mergers since $z \sim 1.2$. The weak evolution predicted for the galaxy merger rate in an hierarchical model universe shows that the mass assembly evolution of galaxies through mergers does not follow the rapid evolution of the halo merger rate obtained in previous studies.
\end{abstract}


\section{Introduction}

The role played by galaxy mergers is of paramount importance for galaxy formation and evolution. Since the seminal numerical work by \citet{toomres72}, galaxy mergers are thought to be relevant mechanisms in the build up of the spheroidal component of galaxies. Recent numerical works also argue that gas-rich mergers are able to form disk galaxies \citep{robertson04,governato07}. In the local Universe, observations of remnants of galaxy mergers strengthen their effects on galaxy morphologies and on triggering star formation in merging galaxies, unveiling their power on driving galaxy evolution \citep[e.g.][]{jog06,schweizer07}. Actually, in a recent work \citet*{mateus08} show that galaxy mergers are very correlated to the star formation history of galaxies, driving the environmental dependence in galaxy formation by regulating the star formation process. 

The importance of galaxy mergers is highlighted when we consider the hierarchical build up of cold dark matter (CDM) structures according to the standard $\Lambda$CDM cosmology, which forms the basis of the subsequent processes related to galaxy formation and evolution. In this picture, the structure growth is driven by the hierarchical accretion and merging of dark matter haloes, with galaxies forming inside these structures following the radiative cooling of baryons and the condensation of the gas through dissipative cooling \citep{white_rees78}. It follows that the rate at which galaxy mergers evolve is a direct consequence of the hierarchical growth of structures in any galaxy formation model based on this framework. Predicting this evolution from the models and comparing it to observations of galaxy mergers at different lookback times represent, therefore, a key test to determine whether and how galaxies form by mergers or whether the models give us the correct predictions. 

On the observational side, the galaxy merger counting is a challenging effort. Two main approaches are used to measure the evolution of galaxy mergers at low and high-redshift. The first one is based on the counting of close galaxy pairs, which provides reliable estimates of major mergers out to $z \sim 1$ \citep[e.g.][]{patton97, lefevre00, bundy04, lin04, karta07}, despite of the selection effects and biases inherent to this method \citep[e.g.][]{patton00}. Other way to counting galaxy mergers is through detecting direct signatures of galaxy interactions, as visual disturbed morphologies or asymmetries detected by automated procedures \citep[e.g.][]{conselice03,depropris07,lotz08}.

In this Letter, we examine the galaxy merger rate evolution predicted by galaxy formation models and its comparison with observational measurements of merger rates taken from the literature. We use the public data available by the Millennium Simulation \citep{springel05} to derive galaxy merger properties predicted by two semi-analytical galaxy formation models implemented on the simulation and described by  \citet{bower06} and \citet{delucia_blaizot07}. 

We present a brief description of the Millennium Simulation and the semi-analytical models used in this work in Section~\ref{model}. In Section~\ref{description} we describe the methodology used to estimate galaxy mergers from the simulation outputs. The main results of this work are presented in Section~\ref{results} and our conclusions are summarized on Section~\ref{summary}.

\section{The model universes}\label{model}

In this section we briefly describe the model universes used in this work based on one of the largest cosmological simulation of structure growth carried out so far, the Millennium Simulation, and on two state-of-the-art semi-analytical models of galaxy formation developed by independent groups.

The Millennium Simulation of dark matter structure growth, recently carried out by the Virgo Consortium, follows the evolution of $N = 2160^3$ particles of mass $8.6 \times 10^8 h^{-1}$~\msun, within a co-moving box of size $500 h^{-1}$~Mpc on a side, from redshift $z = 127$ to the present \citep{springel05}. The adopted cosmological model is the concordance $\Lambda$CDM model with parameters $\Omega_m = 0.25$, $\Omega_b=0.045$, $h=0.73$, $\Omega_\Lambda=0.75$, $n=1$ and $\sigma_8=0.9$, where the Hubble constant is parametrized as $H_0 = 100 h$~km~s$^{-1}$~Mpc$^{-1}$. Haloes and subhaloes in 64 output snapshots were identified using the SUBFIND algorithm described in \citet{springel01}, and merger trees were then constructed that describe how haloes grow as the model universe evolves. These merger trees form the basic input needed by the semi-analytic models described in the following.

We explore the galaxy catalogues obtained by two semi-analytical galaxy formation models implemented on the Millennium Simulation. We use the results obtained by the model presented and discussed by \citet{delucia_blaizot07}, which is a modified version of the models used in \citet{croton06} and \citet{delucia06}. The galaxy catalogues generated by this model have been made public on the Millennium download site at the German Virtual Observatory\footnote{http://www.g-vo.org/Millennium}. We also use the model described  in detail by \citet{bower06}. It is an extension of the GALFORM model implemented by \citet{cole00} and \citet{benson03}.  The catalogues generated by this model are publicly available on the Millennium download site at the University of Durham.\footnote{http://galaxy-catalogue.dur.ac.uk:8080/Millennium}

These semi-analytical galaxy formation models have significant differences in the general procedure to form  galaxies from the dark matter haloes. First of all, many aspects of the baryonic physics involved in the galaxy formation process are treated in a completely different manner, mainly those related to the energy feedback from active galaxy nuclei to regulate the star formation in galaxies. In addition, the construction of the merger trees from the dark matter halo distribution, which form the basis of the semi-analytical approach integrated on the numerical simulations, has been done in independent ways. 

\section{Galaxy merger rates}\label{description}

We derive the predicted evolution of the galaxy merger rate from the semi-analytical galaxy formation models implemented on the Millennium Simulation. We start selecting all $z=0$ galaxies with rest-frame absolute magnitudes $M_B \leq -18$, resulting in two base catalogues containing about $3.33\times10^6$ galaxies from the \citet{bower06} model and about $4.35\times10^6$ galaxies from the \citet{delucia_blaizot07} model. For all galaxies in these catalogues, we recovered their complete merger trees by using the post-processing facilities carried out on the Millennium databases.

The first step to derive the model predictions for the fraction of galaxies undergoing mergers is to identify the progenitors of each galaxy selected from the merger trees at each time-step $t$ of the simulation.  We then define that a galaxy has experienced a merger if it has at least one pair of progenitors at the previous time-step. In other words, we are able to select galaxy pairs at a given $t$ that will merge to form a single galaxy in the immediately posterior time-step. This description is particularly interesting if one wants to compare the results from the simulation with observational estimates of merger fractions based on pair count statistics \citep[e.g.][]{patton02,lin04,karta07}. In our analysis, the progenitor pair of a merge-product galaxy is composed by its main progenitor (defined as the progenitor with the largest stellar mass), selected to be brighter than a $B$-band absolute magnitude limit, $M_B^{lim}$, and an ordinary progenitor with  $M_B \le -18$. In our analysis, we considered two magnitude limits for the main progenitors, $M_B^{lim} = -19.5$ and $M_B^{lim} = -20$, in order to inspect how this choice affects the derived pair fractions. We have also adopted an additional approach to select galaxy pairs by considering only the ordinary progenitors with stellar masses larger than $1/4$, $1/6$, and $1/10$ of the main progenitor stellar mass, independently of their luminosity. In such analysis, we selected only the main progenitors with $M_B < -20$. This procedure allows us to investigate how the estimated pair fractions depend on the mass ratio of the pair components.

We define the pair fraction as the number of galaxies in the progenitor pairs (main progenitors plus companions), $N_p(z)$, divided by the total number of main progenitors brighter than $M_B^{lim}$,  $N_t(z)$, at each time-step of the simulation: $f_p(z) = N_p(z)/N_t(z)$. The merge fraction is directly derived from the pair fraction by considering only the total number of progenitors pairs that will merge, that is, $f_{mrg}(z) \simeq 0.5f_p(z)$. We can directly relate this predicted merger fraction to observations of close galaxy pairs which are frequently used to derive observed merger fractions. From observations of physically close galaxy pairs one can determine the number of close companions per galaxy, $N_c$ , which is identical to the pair fraction in a volume-limited sample with no triples or higher order N-tuples. Following the expression adopted by \citet{lin04}, the galaxy merger rate per Gyr estimated through pair count statistics is given by 
\begin{equation}\label{eq:rate}
R_{mrg} = (0.5+G) N_c(z) P_{mrg} T_{mrg}^{-1}
\end{equation}
where $N_c(z)$ is the average number of  companions for galaxies within the observed magnitude range, $P_{mrg}$ denotes the probability that galaxies in close pairs will merge (typically assumed to be 0.5), and $T_{mrg}$ is the time-scale for physically associated pairs to merge. The factor $0.5$ is to convert the number of galaxies into the number of merger events (assuming only pairs, ignoring triples and higher order N-tuples). The physical galaxy pairs are expected to merge on the time-scale $T_{mrg}$ which depends on the intrinsic characteristics of the pairs, such as the relative mass ratio, orbit parameters, and on the structure of the merging galaxies, thus varying for each pair \citep[e.g.][]{conselice06}. In general, one assumes that $T_{mrg} \sim 0.5$~Gyr, a value suggested by simplified models based on the dynamical friction time-scale and also predicted by N-body simulations \citep[e.g.][]{mihos95, patton00, conselice06}. In equation (\ref{eq:rate}) we also include an additional parameter, $G$, introduced by \citet{lin04} to account for differences in sample selections when comparing merger rates estimated by pair counts and by morphological approaches. Here we will adopt the value of $G = 1.24$ determined by \citet{lin04}.

In the case of the simulation, a merger between  one or more ordinary progenitors and the main progenitor results in the formation of a single galaxy at a posterior time-step. Thus, we can evaluate a galaxy merger rate (number of galaxy mergers per Gyr), $\Gamma_{mrg}(z)$, by dividing the merger fraction at each $z$ by the time interval between the two time-steps associated to the progenitors and to the new formed galaxy: $\Gamma_{mrg}(z) = f_{mrg}(z) \delta t^{-1}$.  In the redshift range $0<z<1$, the mean value of $\delta t$ is $0.34$~Gyr, implying that the progenitor pairs of a given galaxy can be considered as physical pairs. Thus, the predicted galaxy merger rate can be easily compared with that derived from observational data, $R_{mrg}$, based on pair count statistics.

\section{Results}\label{results}

\begin{figure}
\centering{\includegraphics[width=\columnwidth]{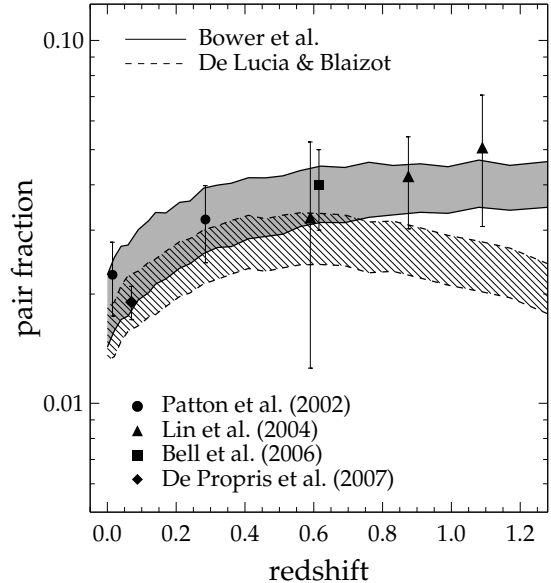}}
\caption{Evolution of the pair fractions predicted by the semi-analytical galaxy formation models of \citet[solid lines]{bower06} and \citet[dashed lines]{delucia_blaizot07} and the observational data taken from different sources for galaxies in the redshift range $0 < z < 1.2$. The lines correspond to model predictions considering $M_B^{lim} = -19.5$ (bottom lines) and $M_B^{lim} = -20$ (top lines) to select the main progenitors, as described in Section~\ref{description}.}
\label{fig:1}
\end{figure}

In  Fig.~\ref{fig:1}, we show the predicted evolution of pair fractions derived from galaxy formation models, $f_p(z)$, and a compilation of observational results in the redshift range $0 < z < 1.2$. For the simulation, we show results of pair fractions obtained assuming magnitude limits to select the main progenitors of $M_B^{lim} = -19.5$ (bottom lines) and $M_B^{lim} = -20$ (top lines). The observational data points shown in this figure were obtained from estimates using galaxy pair counts by \citet{patton02}, \citet{lin04}, \citet{bell06}, and \citet{depropris07}. For $z < 0.3$, we show the close pair fractions obtained by Patton et al. from counting close dynamical pairs of the CNOC2 survey, with luminosities in the range  $-22 \la M_B \la -19$, and the result obtained by De Propris et al. for a galaxy sample with $-21 < M_B -5\log h< -18$ and $0.010 < z < 0.123$ drawn from the Millennium Galaxy Catalogue. At higher redshifts, we show the mean value of the close pair fraction obtained by Bell et al. after analysing  $0.4 < z < 0.8$ galaxies with $M_B < -20$ from the COMBO-17 survey and the pair fractions obtained by Lin et al. from analysis of galaxies with luminosities $-22 \la M_B \la -20$ and redshift range $0.5 < z < 1.2$ from the DEEP2 redshift survey. It is worth noting that we have adjusted the values of Lin et al. downward by 65 per cent to account for projection within galaxy groups, as suggested by \citet{bell06}.

There is a good match between the evolution of the pair fractions predicted by the \citet{bower06} model and that inferred by the observational data, considering the magnitude limits adopted to select the main progenitors. The model by \citet{delucia_blaizot07} underpredicts the pair fractions at $z > 0.5$, showing a decreasing evolution which is clearly inconsistent with the behaviour expected from the observations. The evolution of the pair fractions derived from the Bower et al. model is strong at $z \la 0.3$, and becomes weak at higher redshifts, independent of the value of $M_B^{lim}$ adopted to select galaxy pairs. The amplitude of the pair fractions is highly dependent on the value of $M_B^{lim}$, being higher for pairs containing brighter galaxies. This trend is associated with the strong dependence of the galaxy growth through mergers on stellar mass \citep{maller06,guo08}. We have also compared our results with the recent measurements of close pair fractions obtained by \citet{karta07}, who claims for a strong evolution of the pair fractions since $z \sim 1.2$ after analysing the galaxy population in the COSMOS field. They adopted photometric redshifts to select galaxy pairs in the redshift range $0.1 < z < 1.2$, with a magnitude limit $M_V < -19.8$. We found a good agreement in the pair fractions for $z \la 1$. For higher redshifts, there is an excess in the pair fractions derived by \citet{karta07}, inconsistent with the weak evolution of $f_p(z)$ predicted by the Bower et al. model. This discrepancy can be due to the large errors in the photometric redshifts used to select the pairs and the consequent projection effects, which become more important at higher redshifts. We hope that a future spectroscopic follow-up of their data and other pair fraction estimates from high-$z$ surveys will help to clarify this issue.

\begin{figure}
\centering{\includegraphics[width=\columnwidth]{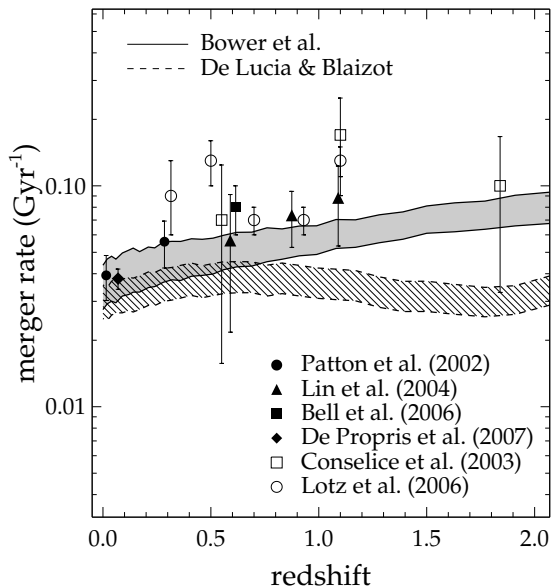}}
\caption{Evolution of the galaxy merger rate predicted by the semi-analytical galaxy formation models of \citet[solid line]{bower06} and \citet[dashed line]{delucia_blaizot07} and the observational data taken from different sources for galaxies in the redshift range $0 < z < 2$ (see Section~\ref{results} for detailed description). The lines are as in Fig.~\ref{fig:1}.}
\label{fig:2}
\end{figure}

In Fig.~\ref{fig:2}, we show the evolution of the merger rates predicted by the models, $\Gamma_{mrg}(z)$, and the merger rates derived from the observations. Galaxy merger rates are estimated from the observational data considering the time-scale $T_{mrg} = 0.5$~Gyr for the measurements from pair counts, assuming by simplicity that this value is characteristic of the merging time-scale for close pairs \citep[e.g.][]{patton00}, as discussed in Section~\ref{description}. For this comparison, we also added the merger rates estimated from merger analysis based on galaxies showing morphological distortions. This approach is significantly distinct to the pair counts analysis, since the mergers are observed after their occurrence (see \citealt{depropris07} for a discussion between these two methods). We show the results obtained by \citet{conselice03} from a morphological analysis of galaxies in the Hubble Deep Field. They have detected ongoing major mergers by measuring an asymmetry parameter from galaxy images. We also show the recent measurements of morphologically identified galaxy mergers obtained by \citet{lotz08} from the analysis of galaxies brighter than $M_B = -20.5$ in the HST survey of the Extended Groth Strip. Here we adopted a time-scale of $1$~Gyr to derive the merger rates from these morphological studies. 

As in Fig.~\ref{fig:1}, there is a good agreement between the observational data points and the Bower et al. model predictions, considering the error bars of the measurements and all the uncertainties involved in deriving the observed merger rates (e.g. time-scales). The evolution of merger fractions derived from observational data sets is usually fitted with a function $f \propto (1+z)^m$, where the value of $m$ varies from about 0 to 4 \citep[e.g.][]{lefevre00, carlberg00, conselice03, lin04, lotz08}. In our analysis, the fitting of the Bower et al. model predictions with the same function, considering only the points in the redshift range $0 < z < 2$, gives:
\begin{eqnarray}\label{eq:2}
\Gamma_{mrg}(z) = 0.029(1+z)^{0.767} \quad \textrm{for} \quad M_B^{lim} = -19.5 \textrm{,}&\textrm{and}\nonumber
\\
\Gamma_{mrg}(z) = 0.046(1+z)^{0.596} \quad \textrm{for} \quad M_B^{lim} = -20.0\textrm{.}&
\end{eqnarray}
The $m \sim 0.6-0.8$ value obtained for this model is consistent with recent results obtained by different approaches, which gives support for roughly a flat evolution of the merger rate \citep{bundy04,lin04,lotz08}. The model by \citet{delucia_blaizot07} predicts a constant merger rate evolution, visibly at odds with the observations and the Bower et al. model results. 

\begin{figure}
\centering{\includegraphics[width=\columnwidth]{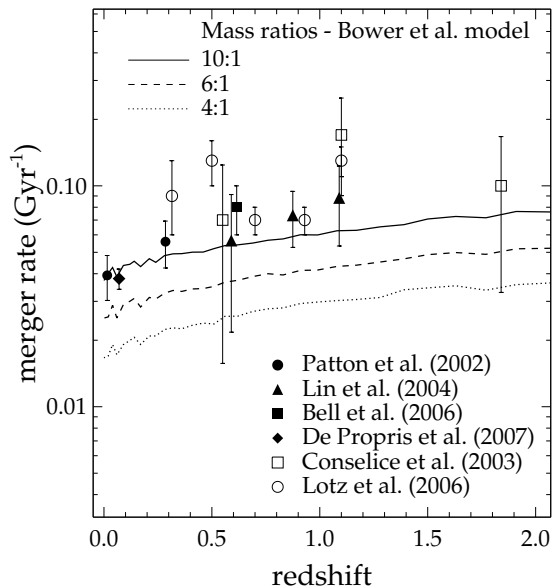}}
\caption{Evolution of the galaxy merger rate predicted by the semi-analytical galaxy formation model of \citet{bower06} considering mass ratios of 4:1, 6:1, and 10:1 to select galaxy pairs, as described in Section~\ref{description}.} 
\label{fig:3}
\end{figure}

In Fig.~\ref{fig:3} we investigate the dependence of the merger rate on the mass ratio of the pair components. We show results for the Bower et al. model from which we selected galaxy pairs with mass ratios smaller than 4:1, 6:1, and 10:1, as described in Section~\ref{description}. The amplitude of the merger rate derived from the model depends strongly on the mass ratio adopted to select the pairs, and it is compatible with the observed one only when we consider a mass ratio threshold of 10:1. The fit of the results with the same function as in equation (\ref{eq:2}) gives $m = 0.59$ for merger rates derived from pairs with mass ratios smaller than 10:1, $0.63$ for mass ratios of 6:1, and $0.68$ for mass ratios of 4:1. These results are consistent with a roughly flat evolution of the merger rate independently of the mass ratio threshold adopted.

\section{Summary and Conclusions}\label{summary}

In this Letter, we have explored galaxy mergers in model universes drawn from semi-analytical galaxy formation models implemented on the Millennium Simulation of structure growth. From the public galaxy catalogues and merger trees generated by the \citet{bower06} and \citet{delucia_blaizot07} models we retrieved the pair fractions and the galaxy merger rate (number of galaxy mergers per Gyr), and compared their redshift evolution with observational results in the redshift range $0 < z < 2$. 

The results from our study indicate that the evolution of the galaxy merger rate has been discrete since $z \sim 2$, following an exponential function given by $\Gamma_{mrg} \propto (1+z)^{m}$, with $m$ ranging from $0.6$ to $0.8$ depending on the luminosity and mass ratios of the merging galaxies. We find that this behaviour is true only for the Bower et al. model, since the model by \citet{delucia_blaizot07} predicts a null merger rate evolution ($m \sim 0$), at odds with the observations and the Bower et al. trends. The results for the \citet{delucia_blaizot07} model can be related to the study carried out by \citet{guo08} on the galaxy growth through mergers and star formation based on it, where they have found a nearly constant growth by mergers in the range $0 < z < 2$, independent of the galaxy stellar mass.

The weak evolution of the galaxy merger rate found in our analysis is consistent with recent observational results obtained by different approaches which give support for a flat evolution of the galaxy merger rate, with $m \sim 0.5-1$ \citep{bundy04,lin04,lotz08}. Previous studies have shown that in the hierarchical framework the merger rate of  cold dark matter haloes increases rapidly with redshift also as $(1 + z)^m$, but with the exponent $m \sim 2-3$ \citep[e.g.][]{gottlober01,fakhouri07}. However, as discussed by \citet{berrier06}, multiple galaxies may occupy the same parent dark matter halo, resulting that the predicted galaxy merger rate should evolve slower than the halo merger rate. Actually, in the present work we have confirmed this tendency. The weak evolution of the galaxy merger rate predicted by the galaxy formation models shows that the mass assembly evolution of galaxies through mergers does not follow the rapid evolution of the halo merger rate obtained in previous studies.

\section*{Acknowledgments}
The author thanks the hospitality of the Institut de Ci\'encies de l'Espai (IEEC, Barcelona), where the present work was developed. This work was supported by the European Commission's ALFA-II programme through its funding of the Latin-american European Network for Astrophysics and Cosmology (LENAC).

\bsp

\label{lastpage}


\begin{thebibliography}{}

\bibitem[\protect\citeauthoryear{Bell et al.}{2006}]{bell06} Bell E.~F., Phleps S., Somerville R.~S., Wolf C., Borch A., Meisenheimer K., 2006, ApJ, 652, 270
\bibitem[\protect\citeauthoryear{Benson et al.}{2003}]{benson03} Benson A.~J., Bower R.~G., Frenk C.~S., Lacey C.~G., Baugh C.~M., Cole S., 2003, ApJ, 599, 38
\bibitem[\protect\citeauthoryear{Berrier et al.}{2006}]{berrier06} Berrier J.~C., Bullock J.~S., Barton E.~J., Guenther H.~D., Zentner A.~R., Wechsler R.~H., 2006, ApJ, 652, 56 
\bibitem[\protect\citeauthoryear{Bower et al.}{2006}]{bower06} Bower R.~G., Benson A.~J., Malbon R., Helly J.~C., Frenk C.~S., Baugh C.~M., Cole S., Lacey C.~G., 2006, MNRAS, 370, 645
\bibitem[\protect\citeauthoryear{Bundy et al.}{2004}]{bundy04} Bundy K., Fukugita M., Ellis R.~S., Kodama T., Conselice C.~J., 2004, ApJ, 601, L123 
\bibitem[\protect\citeauthoryear{Carlberg et al.}{2000}]{carlberg00} Carlberg R. G., et al., 2000, ApJ, 532, L1
\bibitem[\protect\citeauthoryear{Cole et al.}{2000}]{cole00} Cole S., Lacey C.~G., Baugh C.~M., Frenk C.~S., 2000, MNRAS, 319, 168
\bibitem[\protect\citeauthoryear{Conselice et al.}{2003}]{conselice03} Conselice C.~J., Bershady M.~A., Dickinson M., Papovich C., 2003, AJ, 126, 1183
\bibitem[\protect\citeauthoryear{Conselice}{2006}]{conselice06} Conselice C.~J., 2006, ApJ, 638, 686 
\bibitem[\protect\citeauthoryear{Croton et al.}{2006}]{croton06} Croton D.~J., et al., 2006, MNRAS, 365,11
\bibitem[\protect\citeauthoryear{De Lucia et al.}{2006}]{delucia06} De Lucia G., Springel V., White S.~D.~M., Croton D., Kauffmann G., 2006, MNRAS, 366, 499
\bibitem[\protect\citeauthoryear{De Lucia \& Blaizot}{2007}]{delucia_blaizot07} De Lucia G., Blaizot J., 2007, MNRAS, 375, 2
\bibitem[\protect\citeauthoryear{De Propris et al.}{2007}]{depropris07} De Propris R., Conselice C.~J., Liske J., Driver S.~P., Patton D.~R., Graham A.~W., Allen P.~D., 2007, ApJ, 666, 212 
\bibitem[\protect\citeauthoryear{Fakhouri \& Ma}{2007}]{fakhouri07} Fakhouri O., Ma C.-P., 2007, pre-print, arXiv:astro-ph/0710.4567
\bibitem[\protect\citeauthoryear{Gottl\"ober, Klypin \& Kravtsov.}{2001}]{gottlober01} Gottl\"ober S., Klypin A., Kravtsov  A. V., 2001, ApJ, 546, 223
\bibitem[\protect\citeauthoryear{Governato et al.}{2007}]{governato07} Governato F., Willman B., Mayer L., Brooks A., Stinson G., Valenzuela O., Wadsley J., Quinn T., 2007, MNRAS, 374, 1479
\bibitem[\protect\citeauthoryear{Guo \& White}{2008}]{guo08} Guo Q., White S.~D.~M., 2008, MNRAS, 384, 2
\bibitem[\protect\citeauthoryear{Jog \& Maybhate}{2006}]{jog06} Jog C.~J., Maybhate A., 2006, MNRAS, 370, 891 
\bibitem[\protect\citeauthoryear{Kartaltepe et al.}{2007}]{karta07} Kartaltepe J.~S., et al., 2007, ApJS, 172, 320 
\bibitem[\protect\citeauthoryear{Le F{\`e}vre et al.}{2000}]{lefevre00} Le F{\`e}vre O., et al., 2000, MNRAS, 311, 565
\bibitem[\protect\citeauthoryear{Lemson et al.}{2006}]{lemson06} Lemson G., The Virgo Consortium, 2006, pre-print, arXiv:astro-ph/0608019 
\bibitem[\protect\citeauthoryear{Lin et al.}{2004}]{lin04} Lin L., et al., 2004, ApJ, 617, L9
\bibitem[\protect\citeauthoryear{Lotz et al.}{2008}]{lotz08} Lotz J.~M., et al., 2008, ApJ, 672, 177
\bibitem[\protect\citeauthoryear{Maller et al.}{2006}]{maller06} Maller A.~H., Katz N., Kere{\v s} D., Dav{\'e} R., Weinberg D.~H., 2006, ApJ, 647, 763
\bibitem[\protect\citeauthoryear{Mateus, Jimenez, \& Gazta\~naga}{Mateus et al.}{2008}]{mateus08} Mateus A., Jimenez R., \& Gazta\~naga E., 2008, pre-print, arXiv:astro-ph/0801.3282
\bibitem[\protect\citeauthoryear{Mihos}{1995}]{mihos95} Mihos J.~C., 1995, ApJ, 438, L75 
\bibitem[\protect\citeauthoryear{Patton et al.}{1997}]{patton97} Patton D.~R., Pritchet C.~J., Yee H.~K.~C., Ellingson E., Carlberg R.~G., 1997, ApJ, 475, 29 
\bibitem[\protect\citeauthoryear{Patton et al.}{2000}]{patton00} Patton D.~R., Carlberg R.~G., Marzke R.~O., Pritchet C.~J., da Costa L.~N., Pellegrini P.~S., 2000, ApJ, 536, 153 
\bibitem[\protect\citeauthoryear{Patton et al.}{2002}]{patton02} Patton D.~R., et al., 2002, ApJ, 565, 208
\bibitem[\protect\citeauthoryear{Robertson et al.}{2004}]{robertson04} Robertson B., Yoshida N., Springel V., Hernquist L., 2004, ApJ, 606, 32
\bibitem[\protect\citeauthoryear{Schweizer \& Seitzer}{2007}]{schweizer07} Schweizer F., Seitzer P., 2007, AJ, 133, 2132 
\bibitem[\protect\citeauthoryear{Springel et al.}{2001}]{springel01} Springel V., White S.~D.~M., Tormen G., Kauffmann G., 2001, MNRAS, 328, 726
\bibitem[\protect\citeauthoryear{Springel et al.}{2005}]{springel05} Springel V., et al., 2005, Nature, 435, 629
\bibitem[\protect\citeauthoryear{Toomre \& Toomre}{1972}]{toomres72} Toomre A., Toomre J., 1972, ApJ, 178, 623
\bibitem[\protect\citeauthoryear{White \& Rees}{1978}]{white_rees78} White S.~D.~M., Rees M.~J., 1978, MNRAS, 183, 341

\end{thebibliography}
\end{document}